\documentclass[prb,aps,showpacs,showkeys,groupedaddress,superscriptaddress,twocolumn]{revtex4-2}

\usepackage[utf8x]{inputenc}
\usepackage{color}
\usepackage{bbm} 
\usepackage{amsfonts,amsmath,amssymb,stmaryrd}
\usepackage{esint}
\usepackage{graphicx}
\usepackage{subfigure}  
\usepackage{hyperref}
\usepackage{epsfig}
\usepackage{mathrsfs}
\usepackage{verbatim}
\usepackage{centernot}

\renewcommand{\c}{\hat{c}}

\newcommand{\cd}{\hat{c}^\dagger}

\usepackage{array}

\usepackage{bm}	
\renewcommand{\vec}[1]{\bm{#1}}

\begin{document}
\title{Chern number and Berry curvature for Gaussian mixed states of fermions}

\author{Lukas Wawer}
\affiliation{Department of Physics and Research Center OPTIMAS, University of Kaiserslautern, 67663 Kaiserslautern, Germany}
\author{Michael Fleischhauer}
\affiliation{Department of Physics and Research Center OPTIMAS, University of Kaiserslautern, 67663 Kaiserslautern, Germany}


\begin{abstract}
We generalize the concept of topological invariants for mixed states  based on the ensemble geometric phase (EGP) 
to two-dimensional bandstructures. In contrast to the geometric Uhlmann phase for density matrices 
the EGP leads to a proper Chern number for Gaussian, finite-temperature or non-equilibrium steady states. 
The Chern number can be expressed as an integral of the ground-state Berry curvature 
of a fictitious lattice Hamiltonian, constructed from single-particle correlations.
For the Chern number to be non-zero the fictitious Hamiltonian has to break time-reversal symmetry. 
\end{abstract}
\pacs{}

\date{\today}
\maketitle


\section{Introduction}

Transitions between different phases of quantum matter are  characterized either by a spontaneous breaking of symmetries or by changes of the topology of the many-body ground state \cite{Xiao-RMP-2010,Hazan-Kane-RMP-2010,Wen-RMP-2017}. 
Topologically different phases can be distinguished by invariants, which identify global properties of the system. The existence of these integer quantum numbers is also the origin of the robustness of characteristic features 
such as edge states and currents or quantized bulk transport  \cite{Klitzing-PRL-1980,TKNN-PRL-1982,Tsui-PRL-1982,Laughlin-PRL-1983,Arovas-PRL-1984}.
Typical topological invariants are based on geometric phases such as the Berry or Zak phase \cite{Xiao-RMP-2010}
which characterize the parallel transport of the many-body ground-state 
upon cyclic changes of system parameters.
They are thus applicable only to pure states. 
Several attempts have been made over the last years to extend the concept of topological invariants to mixed states of non-interacting fermions \cite{Avron-NJP-2011,Diehl-NatPhys-2011,Bardyn-NJP-2013,Viyuela-PRL-2014,Huang-PRL-2014,Viyuela-PRL-2014b,Nieuwenburg-PRB-2014,Budich-Diehl-PRB-2015,Linzner-PRB-2016,Bardyn-PRX-2018,McGinley-PRL-2018,McGinley-PRB-2019} with the aim of classifying finite-temperature or non-equilibrium steady states.
Although some aspects of topology in open quantum systems can be captured by  
non-Hermitian Hamiltonians \cite{Kawabata-PRX-2019,Zhou-PRB-2019,Ashida-2020}, a proper classification must account for both dissipation and fluctuations and requires the discussion of density matrices.

A possible generalization of Berry's phase to density matrices has been given by Uhlmann \cite{Uhlmann-Rep-Math-Phys-1986}. Based on Uhlmanns construction a mixed-state topological invariant for one-dimensional (1D) systems was defined in \cite{Viyuela-PRL-2014}
as the winding number of the Uhlmann phase upon cyclic parameter changes. 
For 1D lattice systems in a gapped \emph{ground} state, the Uhlmann phase is identical to the Zak phase. In this limit its winding can thus be expressed as an integral of a Berry curvature over a 2D torus (lattice momentum and time) which is the well-known first Chern number. For mixed states the existence of a proper Berry connection is in general however not guaranteed. As shown in \cite{Budich-Diehl-PRB-2015}
the approach based on the Uhlmann phase fails when applied to two dimensions \cite{Huang-PRL-2014,Viyuela-PRL-2014b}.
 The windings of the Uhlmann phases in $x$ and $y$ direction, $\int \mathrm{d}k_x \partial_{k_x}\phi^U_y(k_x)$ and $\int \mathrm{d}k_y \partial_{k_y}\phi^U_x(k_y)$, are for some parameters not the same, demonstrating that no proper Berry connection exists.
 
 Recently, we have shown that a generalization of Resta's many-body polarization \cite{Resta-PRL-1998} to mixed states,
 termed ensemble geometric phase (EGP) \cite{Linzner-PRB-2016,Bardyn-PRX-2018}, is an alternative way to define a topological invariant for Gaussian  states of fermions in 1D.  In the thermodynamic limit
 the EGP approaches the Zak phase of the lowest band of a single-particle Bloch Hamiltonian, termed fictitious Hamiltonian, which
 is defined by single-particle correlations and thus 
 contains all properties of the Gaussian mixed state.
 The winding number of the EGP is  a topological invariant characterizing this
 fictitious Hamiltonian.
 The EGP can be detected and a non-trivial topology has direct physical consequences such as quantized particle transport in an auxiliary system weakly coupled to the fermion chain
 \cite{Wawer-2000,Wawer-prep}. It can also be
 extended to 1D models with interactions including systems with fractional topological charges \cite{Unanyan-PRL-2020}. 
 
 As is the case for the Uhlmann phase, the EGP depends on the choice of the momentum direction, when considering two-dimensional lattice models. However, as will be shown here, and in contrast to the Uhlmann case, 
the EGP in any spatial direction can  be expressed as a closed-loop integral over the corresponding component of a \emph{single} Berry connection. The latter describes the ground-state wave function of the fictitious Bloch Hamiltonian. Thus based on the EGP one can define a unique Chern number for mixed states in 2D.
To illustrate this we discuss the asymmetric Qi-Wu-Zhang model, which is a
two-dimensional, topologically non-trivial lattice model with two bands. Due to the asymmetry of the band structure the winding of the Uhlmann phase is different in $x$ and $y$ direction in a certain range of temperatures \cite{Budich-Diehl-PRB-2015}. The EGP winding, on the other hand, is always the same in all directions.

\section{Berry curvature and Zak phase}


To set the stage we start by shortly summarizing the topological classification of fermions in terms of the Berry curvature of Bloch states. To this end we consider insulating many-body states of non-interacting fermions on a two-dimensional lattice described by a number-conserving Hamiltonian.
We set the lattice constant $a=1$, consider a total number of $N^2$ unit cells with periodic boundary conditions in $x$ and $y$ directions and 
use $\hbar =1$. 
The operators $\hat c_{\vec{j},\lambda},\hat c_{\vec{j},\lambda}^\dagger$
describe the annihilation and creation of a fermion in the $\vec{j}$th unit cell. $\vec{j}=(j_x,j_y)$ denotes the $x$ and $y$ coordinates of the unit cell, respectively, and  the index $\lambda\in \{1,\dots,p\}$ labels a possible internal degree of freedom within a unit cell. 
Assuming translational invariance for simplicity, the Hamiltonian can be written in second quantization as
\begin{equation}
H= \sum_{\vec{k}} \sum_{\mu,\nu=1}^p  \tilde{c}_\lambda^\dagger(\vec{k}) \, {\sf h}_{\mu\nu}(\vec{k}) \,  \tilde{c}_\nu(\vec{k}).\label{eq:H}
\end{equation}
Here $\vec{k}=(k_x,k_y)$ is the lattice momentum and ${\sf h}$ is the $p\times p$ single-particle Hamiltonian matrix in momentum space. We assume that ${\sf h}$ has multiple bands, separated by finite gaps, and we consider an insulator, i.e assume that the chemical potential $\mu$ lies within a band gap. 

If ${\sf h}$ breaks time-reversal symmetry, the topological properties of a gapped many-body state can be characterized by the
Berry curvature ${\cal F}^{(n)}(\vec{k})$ of all occupied
Bloch bands $n$
\begin{eqnarray}
    &&{\cal F}^{(n)}(\vec{k}) = \Bigl(\vec \nabla_k \times \vec{A}^{(n)}(\vec{k})\Bigr)_z \\
    &&\enspace = i\biggl(\Bigl\langle \partial_{k_x} u_n(\vec{k})\Bigr\vert\partial_{k_y} u_n(\vec{k})\Bigr\rangle- \Bigl\langle \partial_{k_y} u_n(\vec{k})\Bigr\vert\partial_{k_x} u_n(\vec{k})\Bigr\rangle\biggr).
    \nonumber 
\end{eqnarray}
Here $\vert u_n(\vec{k})\rangle$ is the Bloch function of the $n$th band and $\vec{A}^{(n)}(\vec{k})$ is the Berry connection
$A^{(n)}_j(\vec{k})= i \bigl\langle u_n(\vec{k})\bigr\vert\partial_{k_j} u_n(\vec{k})\bigr\rangle$.
While the Berry curvature is not gauge invariant its
integral over the two-dimensional torus of the Brillouin zone is. It furthermore defines an integer-valued topological invariant of the band, the first Chern number
\begin{eqnarray}
    C= \frac{1}{2\pi} \oiint_\textrm{BZ}\!\! \mathrm{d}\vec{k}\, {\cal F}^{(n)}(\vec{k}).
\end{eqnarray}

The Chern number can also be related to the geometric phase picked up by a Block state $\vert u_n(\vec{k})\rangle$ upon parallel transport in momentum space through the Brillouin zone in either $k_x$ or $k_y$ direction. It can be written 
in terms of the winding of the Zak phases, defined as
\begin{equation}
\phi^\mathrm{Zak}_{x}(k_y) = i \oint_\textrm{BZ} \!\!  \mathrm{d}k_x \, \bigl\langle u(\vec{k})\bigr\vert \partial_{k_x} u(\vec{k})\bigr\rangle = \oint_\textrm{BZ}  \!\! \mathrm{d}k_x \, A_x(\vec{k}).\label{eq:Zak}
\end{equation}
or $\phi^\textrm{Zak}_{y}(k_x)$ respectively. Note that we have dropped the band index for simplicity. 
Most importantly the two Zak phases in $x$ and $y$ direction can be expressed as integrals of the components of a single vector, the Berry connection $\vec{A} =i\langle u(\vec{k}) \vert \vec\nabla_{\vec{k}} u(\vec{k})\rangle$. As a consequence the windings of the two phases
\begin{eqnarray*}
    C_x &=& \frac{1}{2\pi} \oint_\textrm{BZ} \!\! \mathrm{d}k_y\, \frac{\partial}{\partial k_y}\phi^\textrm{Zak}_x(k_y),\\ 
    C_y &=& - \frac{1}{2\pi} \oint_\textrm{BZ} \!\!  \mathrm{d}k_x\,  \frac{\partial}{\partial k_x}\phi^\textrm{Zak}_y(k_x)
\end{eqnarray*}
must be identical and equal to the  Chern number
\begin{equation}
    C_x=C_y=C.\label{eq:CxCy}
\end{equation}
%

\section{Geometric phase for density matrices: Uhlmann construction}

Topological invariants for pure states, such as the Chern number, characterize how a gapped many-body state $\vert \psi\rangle$ changes upon a parallel transport along a closed loop in parameter space. The parallel-transport requirement leads directly to the definition of the Berry phase, or for lattice systems to the
 Zak phase, eq.\eqref{eq:Zak}.
The corresponding Berry connection transforms as a $U(1)$ gauge field according to $\vert u(\vec{k})\rangle\to e^{i\chi(\vec{k})} \vert u(\vec{k})\rangle$: $\vec{A}\to \vec{A}-\vec{\nabla}_{\vec{k}}
\chi(\vec{k})$. 

A generalization of geometric phases to  density matrices $\rho$ has been introduced by Uhlmann 
\cite{Uhlmann-Rep-Math-Phys-1986}, who pointed out that the decomposition $\rho=w\cdot w^\dagger$ of an $n\times n$ density matrix into matrices $w$ contains
a gauge freedom $w\to w \, {\sf U}$, where ${\sf U}$ is  a $U(n)$ unitary matrix. Since $\rho$ is positive semi-definite, $w$ can always be represented as 
\begin{equation}
    w = \sqrt{\rho}\, {\sf U}.
\end{equation}
Let $\rho=\rho(\lambda)$ be a uniquely defined mixed state and let $\lambda\in\{0,\Lambda\}$ parametrize a
 closed loop in parameter space such that $\rho(0)=\rho(\Lambda)$.  Requiring parallel transport of the density matrix in generalization 
of Berry's construction, one can then define a $U(n)$ Uhlmann holonomy
\begin{equation}
    {\cal H}_\mathrm{U} = {\sf U}(\Lambda) {\sf U}^\dagger(0) = {\cal P} e^{-i \int_0^\Lambda \mathrm{d}\lambda \vec{A}_\mathrm{U}}.\label{eq:HU}
\end{equation}
Here ${\cal P}$ denotes path ordering and $\vec{A}_\mathrm{U} = i\partial_\lambda {\sf U}(\lambda) {\sf U}^\dagger(\lambda)$ is the $U(n)$ Uhlmann connection. The $U(n)$ gauge freedom can be reduced to $U(1)$ by performing a trace which leads to
the Uhlmann phase
\begin{equation}
    \phi^\mathrm{U} =\textrm{Im}\ln \textrm{Tr}[ \rho(0) {\cal H}_\mathrm{U}].\label{eq:Uhlmann}
\end{equation}
For (pure) ground states of gapped fermionic  models eq.\eqref{eq:Uhlmann} reduces to the well-known Zak phase, if
$\lambda$ is identified  with the lattice momentum $k$.

In \cite{Viyuela-PRL-2014}, the winding of the Uhlmann phase
\eqref{eq:Uhlmann} upon a cyclic change of an external parameter $t$ has been proposed as a topological invariant to classify one-dimensional lattice models in a mixed state $\rho$
\begin{equation}
    \nu^\mathrm{U} = \frac{1}{2\pi} \int_0^T \!\!\! \mathrm{d}t \, \frac{\partial}{\partial t} \phi^\mathrm{U}(t).
\end{equation}
Here the variable $\lambda$ entering the definition of the Uhlmann phase is the quasi momentum $k$ along the chain and the loop integral  extends over the first Brillouin zone $(\lambda \to k\in[-\pi,\pi))$. While this defines a consistent  invariant 
in 1D, its extension to two spatial dimensions as proposed in \cite{Huang-PRL-2014,Viyuela-PRL-2014b} is problematic. As shown by Budich and Diehl
in \cite{Budich-Diehl-PRB-2015}, the windings of the Uhlmann phase in $x$ or $y$ direction are in general not the same, i.e.
\begin{eqnarray}
C_x^\mathrm{U}&=&\frac{1}{2\pi}\oint_{BZ} \!\! \mathrm{d}k_y\, \frac{\partial}{\partial k_y} \phi^\mathrm{U}_x(k_y) \\
&\ne &-\frac{1}{2\pi}\oint_{BZ}\!\!   \mathrm{d}k_x \, \frac{\partial}{\partial k_x} \phi_y^\mathrm{U}(k_x)=C_y^\mathrm{U}.\nonumber
\end{eqnarray}
As an example they considered  a finite-temperature state of a simple topological two-band model with asymmetric band structure 
\begin{align}
    \hat{\mathcal{H}}(\vec{k}) &= \vec{d}(\vec{k}) \cdot \vec{\sigma} = \sum_{j=1}^3 d_j(\vec{k}) \, \sigma_j,  \\
    \vec{d}(\vec{k})&=\begin{pmatrix}
    \sin(k_x)\\
    3\sin(k_y)\\
    1-\cos(k_x)-\cos(k_y)
    \end{pmatrix},
\end{align}
which is a modification of the Qi-Wu-Zhang model \cite{Qi-Wu-Zhang-PRB-2006}.
Here $\vec{\sigma}$ is the vector of Pauli matrices. Its band-spectrum is shown in Fig.\ref{fig:QWZ-model}.
While at temperatures $T=0$ or $T=\infty$ the windings of the Uhlmann phase in both directions are the same, i.e. $C_y^\mathrm{U}(T=0) =C_x^\mathrm{U}(T=0)=1$
and $C_y^\mathrm{U}(T=\infty) =C_x^\mathrm{U}(T=\infty)=0$, there is a range of temperatures where $C_y^\mathrm{U}(T)\ne C_x^\mathrm{U}(T)$.
This shows that in contrast to the full $U(n)$ Uhlmann holonomy, there is no proper $U(1)$ gauge structure underlying the Uhlmann phase.

\begin{figure}[htb]
	\begin{center}
	\includegraphics[width=0.8\columnwidth]{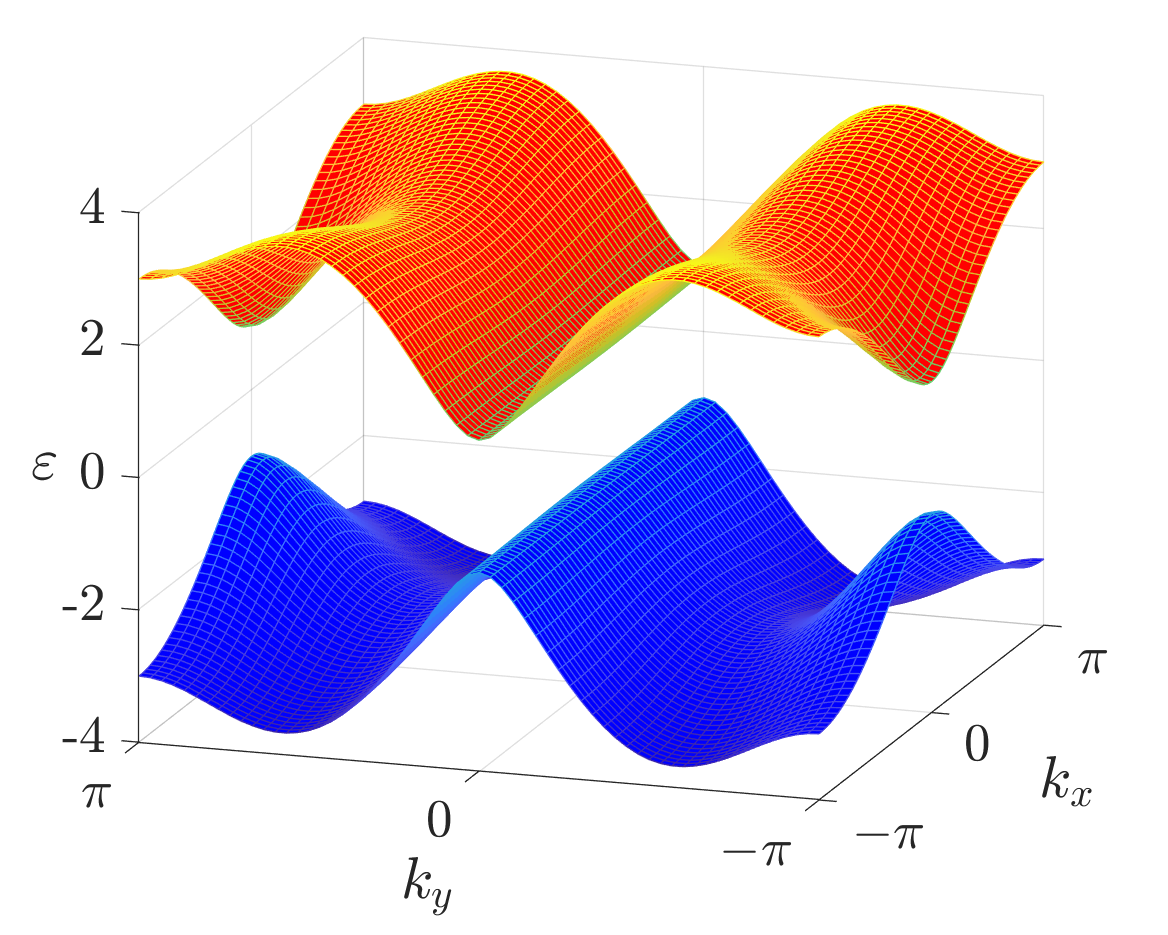}
	\end{center}
	\caption{Spectrum of asymmetric Qi-Wu-Zhang model}
	\label{fig:QWZ-model}
\end{figure}

\section{Many-body polarization and Ensemble Geometric Phase}

In the following we want to show  that in contrast to the Uhlmann phase, the ensemble geometric phase, introduced for one-dimensional lattice models in \cite{Linzner-PRB-2016,Bardyn-PRX-2018}, can be used to define a proper Berry curvature and Chern number for mixed states
of 2D bandstructures.

\subsection{Ground-state Zak phase and many-body polarization}

A physical interpretation of the Zak phase can be picked up from its relation to the many-body polarization of insulating states, which in the form introduced by Resta
reads
\begin{equation}
    P= \frac{1}{2\pi} \textrm{Im}\log \left\langle e^{\frac{2\pi i}{N} \hat X}\right\rangle.
\end{equation}
Here $\hat X = \sum_{j=1}^N \sum_{\lambda=1}^p 
j\hat n_{j;\lambda}$ is the position operator of all particles and the average is 
performed with respect to the insulating many-body ground state $\vert\Psi_0\rangle$. 
Here $\hat n_{j;\lambda} =\hat c_{j;\lambda}^\dagger \hat c_{j;\lambda}$ is the number operator of fermions in the $\lambda$th site of the $j$th unit cell in a periodic system of size $L=pN$. We have disregarded the
relative position of sites within the unit cell. The latter can  straightforwardly be incorporated but does not affect the key properties
of $P$.
One immediately recognizes that the many-body polarization is the phase of a complex number, given by the expectation value of the collective
momentum shift operator $e^{\frac{2\pi i}{N} \hat X}$ devided by $2\pi$. In fact, as shown by King-Smith and Vanderbildt \cite{King-Smith-PRB-1993}, differences of $P$ are strictly related to differences of the Zak phase via
\begin{equation}
\Delta P = \frac{1}{2\pi} \Delta \phi^\textrm{Zak}.
\end{equation}
If the system Hamiltonian is time-dependent with period $T$, the polarization becomes time-dependent as well and its winding upon adiabatic changes in a period $(0,T)$ is the Chern number of a Berry connection on a 2D torus $(k,t)$ of lattice momentum and time
\begin{equation}
C = \frac{i}{2\pi}\int_0^T\!\!  \mathrm{d}t \oint_\textrm{BZ}\!\!  \mathrm{d}k\, \bigl\langle \partial_t u(k)\bigl\vert \partial_k u(k)\bigr\rangle.\label{eq:winding}
\end{equation}
%

\begin{figure}[htb]
	\begin{center}
	\includegraphics[width=0.9\columnwidth]{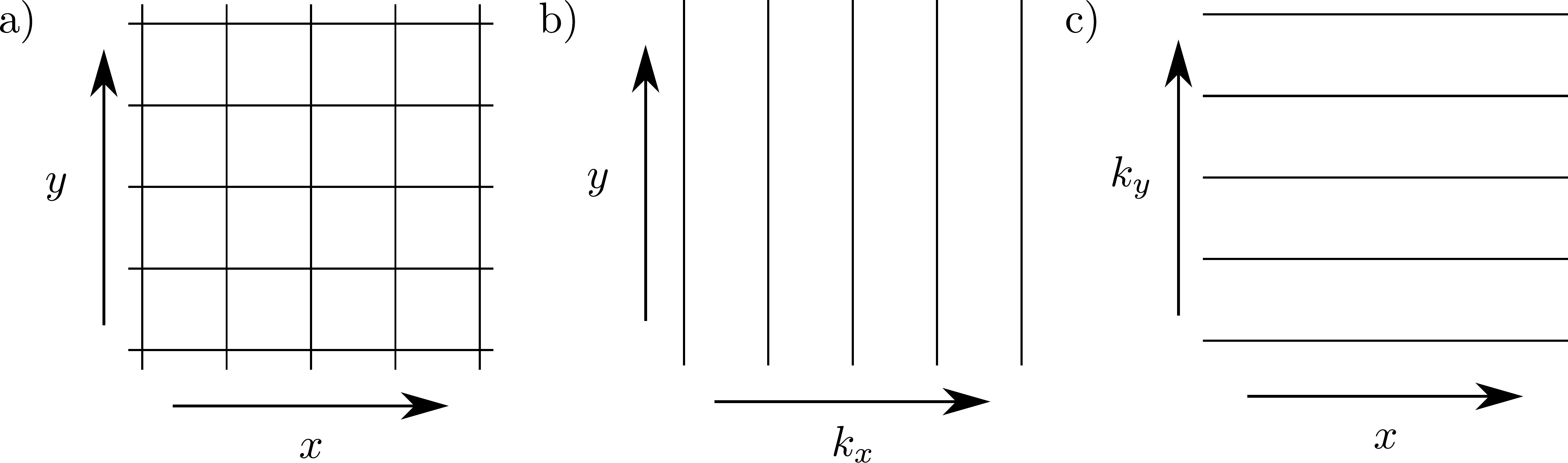}
	\end{center}
	\caption{The Chern number of a two-dimensional, translational invariant lattice model (a) can be represented as the winding of Zak phases of one-dimensional chains. This can be done in two different ways, either by the winding of the Zak phase $\phi_y=\phi_x(k_x)$ upon cyclic changes of the lattice momentum $k_x$ (b), or by the (negative) winding of the Zak phase $\phi_x=\phi_x(k_y)$ upon cyclic changes of $k_y$ (c). }
	\label{fig:decomposition}
\end{figure}

For translationally invariant 2D models one can introduce a polarization vector by mapping the 2D system to a
set of independent 1D chains in either $x$ or $y$ direction, see Fig.\ref{fig:decomposition}:
\begin{eqnarray}
P_x(k_y) &=& \frac{1}{2\pi} \textrm{Im}\log \left\langle e^{\frac{2\pi i}{N} \hat X(k_y)}\right\rangle,\\
P_y(k_x) &=& \frac{1}{2\pi} \textrm{Im}\log \left\langle e^{\frac{2\pi i}{N} \hat Y(k_x)}\right\rangle,
\end{eqnarray}
with $\hat X(k_y) = \sum_{j_x=1}^N \sum_{\lambda=1}^p 
j_x \cd_{j_x,\lambda}(k_y) \c_{j_x,\lambda}(k_y)$, where $\c_{j_x,\lambda}(k_y)$ is the fermion annihilation operator 
in mixed position-momentum space. Similarly $\hat Y(k_x) = \sum_{j_y=1}^N \sum_{\lambda=1}^p 
j_y \cd_{j_y,\lambda}(k_x) \c_{j_y,\lambda}(k_x)$. Applying the King-Smith Vanderbildt relations to the individual components of the polarization vector
\begin{equation}
    \Delta P_x(k_y) = \frac{1}{2\pi} \Delta \phi^\textrm{Zak}_x(k_y),\quad \Delta P_y(k_x) = \frac{1}{2\pi} \Delta \phi^\textrm{Zak}_y(k_x)\nonumber
\end{equation}
and taking into account eq.\eqref{eq:CxCy} shows that there are two equivalent representations of the lattice Chern number in the gapped ground state in terms of polarization components
\begin{equation}
    C_0 = \oint_\textrm{BZ}  \mathrm{d}k_y \, \frac{\partial}{\partial k_y} P_x(k_y)\Bigr\vert_{\Psi_0} = - \oint_\textrm{BZ} \mathrm{d}k_x \, \frac{\partial}{\partial k_x} P_y(k_x)\Bigr\vert_{\Psi_0}.
\end{equation}

\subsection{Ensemble Geometric Phase}

In contrast to the Zak phase, which  has a meaning only for pure states, the many-body polarization can also be evaluated if the system is in a mixed state $\rho$. This then defines the ensemble geometric phase (EGP): 
\begin{equation}
    \phi^\textrm{EGP} = \textrm{Im}\ln\textrm{Tr}\Bigl\{\rho \, e^{\frac{2\pi i}{N} \hat X}\Bigr\}.
\end{equation}
The EGP has been introduced in \cite{Linzner-PRB-2016,Bardyn-PRX-2018} as 
an alternative to the Uhlmann phase for
the definition of topological invariants in 1D lattices. 
Considering the change of the EGP upon an adiabatic, closed loop in parameter space leads to an integer-valued
winding number, which is a topological invariant as long as certain generalized gap conditions are fulfilled \cite{Bardyn-PRX-2018}.
Note that while the EGP is equally applicable to boson systems, mixed states of non-interacting bosons always lead to trivial winding numbers due to the lack of Pauli exclusion \cite{Mink-PRB-2019}.
The EGP can be  measured \cite{Bardyn-PRX-2018} and its non-trivial winding has direct physical implications such as quantized transport in a weakly coupled auxiliary system \cite{Wawer-2000,Wawer-prep}.

\subsection{Berry curvature and Chern number for Gaussian mixed states}

We here consider a special class of mixed states, Gaussian states, which are the analogue of pure many-body eigenstates of non-interacting fermions.
Gaussian states of translationally invariant systems can always be written in the form 
\begin{equation}
\rho = \frac{1}{Z} \exp\left\lbrace-\sum_{\vec k} \mathbf{\tilde{c}}^\dagger({\vec k}) {\sf g}({\vec k}) \mathbf{\tilde{c}}({\vec k})\right\rbrace\label{eq:Gauss}
\end{equation}
and are fully determined by a 
$p\times p$  Hermitian matrix ${\sf g}(\vec{k})$.
Note that we have restricted ourselves for simplicity to systems with particle number conservation.
Here  $\vec{k}=(k_x,k_y)$ is the lattice momentum vector and we  used the abbreviation $ \mathbf{\tilde{c}}(\vec{k})=\bigl(\tilde{c}_1(\vec{k}),\dots, \tilde{c}_p(\vec{k})\bigr)$. 
${\sf g}$ is directly related to the covariance matrix 
of single-particle correlations
\begin{equation}
 {\sf h}^\textrm{fict}_{\mu\nu}(\vec{k}) = \langle  \tilde{c}_\mu^\dagger(\vec{k})  {\tilde{c}}_\nu(\vec{k})\rangle=\frac{1}{2}\left[ 1- \tanh\left(\frac{{\sf g}(\vec{k})}{2}\right)\right]_{\nu,\mu},\label{eq:m}
\end{equation}
which  was termed \emph{fictitious Hamiltonian} in Ref.\cite{Bardyn-NJP-2013}. 
Eq.(\ref{eq:Gauss}) has the form of a grand-canonical
density matrix and indeed for a finite-temperature
state one finds
\begin{equation}
{\sf g}(\vec{k}) = \beta \Bigl( {\sf h}(\vec{k}) -\mu\Bigr).\label{eq:g}
\end{equation}
where ${\sf h}(\vec{k})$ is the (original) single-particle Hamiltonian, $\beta=1/(k_BT)$,  and $\mu$ the chemical potential. 

While all results can be applied to genuine non-equilibrium, Gaussian steady states, we will focus in the following on finite-temperature states. In this context it is important to note that all single-particle Bloch states of the original Hamiltonian ${\sf h}(\vec{k}) $ and that of $ {\sf h}^\textrm{fict}(\vec{k})$ are the same and thus all topological properties carry over. Viewing the fictitious Hamiltonian as the central quantity of Gaussian mixed states also implies that it determines their topological classification following the Altland-Zirnbauer scheme  \cite{Altland-PRB-1997,Schnyder-PRB-2008,Ryu-NJPhys-2010} with 10 distinct symmetry classes.
One further concludes that 
a topological phase transition can take place when the gap of  ${\sf h}^\textrm{fict}$ closes. For thermal states this can happen either if the gap of the original Hamiltonian ${\sf h}$ closes, or if $\beta\to 0$.
Thus the critical temperature of a topological phase transition of the fictitious Hamiltonian is always $T_c=\infty$ irrespective of the details of ${\sf h}$ as long as it remains gapped. This is a crucial difference to the Uhlmann phase, where for several examples a \textit{finite} critical temperature was found \cite{Viyuela-PRL-2014,Viyuela-PRL-2014b}.

If we consider an equilibrium state where the chemical potential is within a band gap of ${\sf h}$, also ${\sf h}^\textrm{fict}$ has an energy gap which is centered around zero ''energy''. 
 It was shown in \cite{Bardyn-PRX-2018} that for 1D bandstructures the EGP then reduces in the thermodynamic limit to the Zak phase of the
 many-body ground state $\vert \Psi_0\rangle$ of the fictitious Hamiltonian via a mechanism termed gauge-reduction:
\begin{equation}
    \phi^\textrm{EGP}(\rho) = \phi^\textrm{Zak}\Bigl(\vert \Psi_0\rangle\langle \Psi_0\vert\Bigr) +{\cal O}(N^{-\alpha}),
\end{equation}
with some $\alpha >0$. Furthermore
since the winding of $\phi^\textrm{EGP}$ as well as that of $\phi^\textrm{Zak}$ upon an adiabatic parameter loop must be a multiple of $2\pi$, the winding numbers for any system size $N$ are equal,$\nu^\textrm{Zak}\vert_{\Psi_0}=\nu^\textrm{EGP}\vert_\rho$, and
\begin{eqnarray}
    \nu^\textrm{EGP}\bigr\vert_\rho &=& \frac{1}{2\pi} \int_0^\Lambda\!\!  \mathrm{d}\lambda \, \frac{\partial}{\partial \lambda} \phi^\textrm{EGP}\\
    &=& \frac{i}{2\pi} \int_0^\Lambda\!\!  \mathrm{d}\lambda \oint_\textrm{BZ}\!\!  \mathrm{d}k \, 
    \biggl(\Bigl\langle \partial_{\lambda} u_f(k)\Bigr\vert\partial_{k} u_f(k)\Bigr\rangle- c.c.\biggr)\nonumber
\end{eqnarray}
 where $\vert u_f(k)\rangle$ are the Bloch states of the negative energy bands of the fictitious Hamiltonian. 
 \begin{equation}
     {\sf h}^\textrm{fict}(k)\,  \vert u_f^{(n)}(k)\rangle = \varepsilon_n(k)\, \vert u_f^{(n)}(k)\rangle.
 \end{equation}
 We note that while for thermal states the Bloch wave-functions of $ {\sf h}^\textrm{fict}$ are just those of the original Hamiltonian ${\sf h}$, the $\vert u_f(k)\rangle$'s have a meaning of their own in the case of a non-equilibrium steady state.

We now argue that the same is true for a finite-temperature state in two spatial dimensions, if the system is translationally invariant, i.e. if a decomposition in independent one-dimensional systems as shown in Fig.\ref{fig:decomposition} is possible. 
In such a case the density matrix at non-zero temperature can be decomposed in two different ways
\begin{eqnarray}
\rho &=& \frac{1}{Z}\prod_{k_x} \exp\left\{-\sum_{q} \hat{\mathbf{c}}^\dagger(k_x,q)\, {\sf g}(k_x,q)\, \hat{\mathbf{c}}(k_x,q)\right\}\label{eq:rho-fact}\\
&=& \frac{1}{Z}\prod_{k_y} \exp\left\{-\sum_{q} \hat{\mathbf{c}}^\dagger(q,k_y)\, {\sf g}(q,k_y)\, \hat{\mathbf{c}}(q,k_y)\right\}.\nonumber
\end{eqnarray}
Then, following the lines of \cite{Bardyn-PRX-2018} the winding of the ensemble geometric phase in $x$ or $y$ direction upon moving $k_y$ or $k_x$ through the Brillouin zone is the same as that of the Zak phase in the ground state of the fictitious Hamiltonian
\begin{eqnarray}
C_x^\textrm{EGP} &=& \frac{1}{2\pi}\oint_\textrm{BZ}\!\!   \mathrm{d}k_y\, \frac{\partial}{\partial k_y} \phi_x^\textrm{EGP}(k_y) \\
&=& 
\frac{i}{2\pi}\oiint_\textrm{BZ} \!\! \mathrm{d}k_x\, \mathrm{d}k_y \biggl(\langle \partial_{k_x} u_f(\vec{k})\vert \partial_{k_y} u_f(\vec{k})\rangle -c.c.\biggr)   \nonumber\\
C_y^\textrm{EGP} &=& -\frac{1}{2\pi}\oint_\textrm{BZ} \!\! \mathrm{d}k_x\, \frac{\partial}{\partial k_x} \phi_y^\textrm{EGP}(k_x) \\
&=& 
-\frac{i}{2\pi}\oiint_\textrm{BZ} \!\! \mathrm{d}k_x\, \mathrm{d}k_y \biggl(\langle \partial_{k_y} u_f(\vec{k})\vert \partial_{k_x} u_f(\vec{k})\rangle - c.c.\biggr).\nonumber
\end{eqnarray}
Obviously both expressions are the same and can be written as an integral of a Berry curvature ${\cal F}^\textrm{EGP}(\vec{k})$ over the two-dimensional Brillouin zone
\begin{eqnarray}
C^\textrm{EGP} &=& C_x^\textrm{EGP} = C_y^\textrm{EGP}\\
&=& \frac{1}{2\pi} \oiint_\textrm{BZ} \!\! \mathrm{d} \vec{k}\, {\cal F}^\textrm{EGP}(\vec{k}).\nonumber
\end{eqnarray}
${\cal F}^\textrm{EGP}(\vec{k})$ is the Berry curvature of the ground state of the fictitious Hamiltonian
\begin{eqnarray}
    &&{\cal F}^\textrm{EGP}(\vec{k}) = \Bigl(\vec \nabla_k \times \vec{A}^\textrm{EGP}(\vec{k})\Bigr)_z \\
    &&\enspace = i\Bigl(\Bigl\langle \partial_{k_x} u_f(\vec{k})\Bigr\vert\partial_{k_y} u_f(\vec{k})\Bigr\rangle- \Bigl\langle \partial_{k_y} u_f(\vec{k})\Bigr\vert\partial_{k_x} u_f(\vec{k})\Bigr\rangle\Bigr).
    \nonumber 
\end{eqnarray}
We conclude that for the two-dimensional generalization of the ensemble geometric phase there exists always 
a proper Berry curvature. The corresponding Chern number can be non-zero only if the fictitious Hamiltonian ${\sf h}^\textrm{fict}(\vec{k})$  breaks time-reversal symmetry. For thermal states this is the case if the original Hamiltonian ${\sf h}(\vec{k})$ breaks time-reversal
symmetry. 

The reduction of topological properties of Gaussian mixed states of fermions in 2D to the ground state of the
fictitious Hamiltonian is fully consistent with the general finding that there are only 10 symmetry classes to classify steady states of open systems rather than 28 as expected, for example, for non-Hermitian Hamiltonians
  \cite{Lieu-PRL-2020,Altland-PRX-2021}.

\begin{figure}[htb]
	\begin{center}
	\includegraphics[width=0.8\columnwidth]{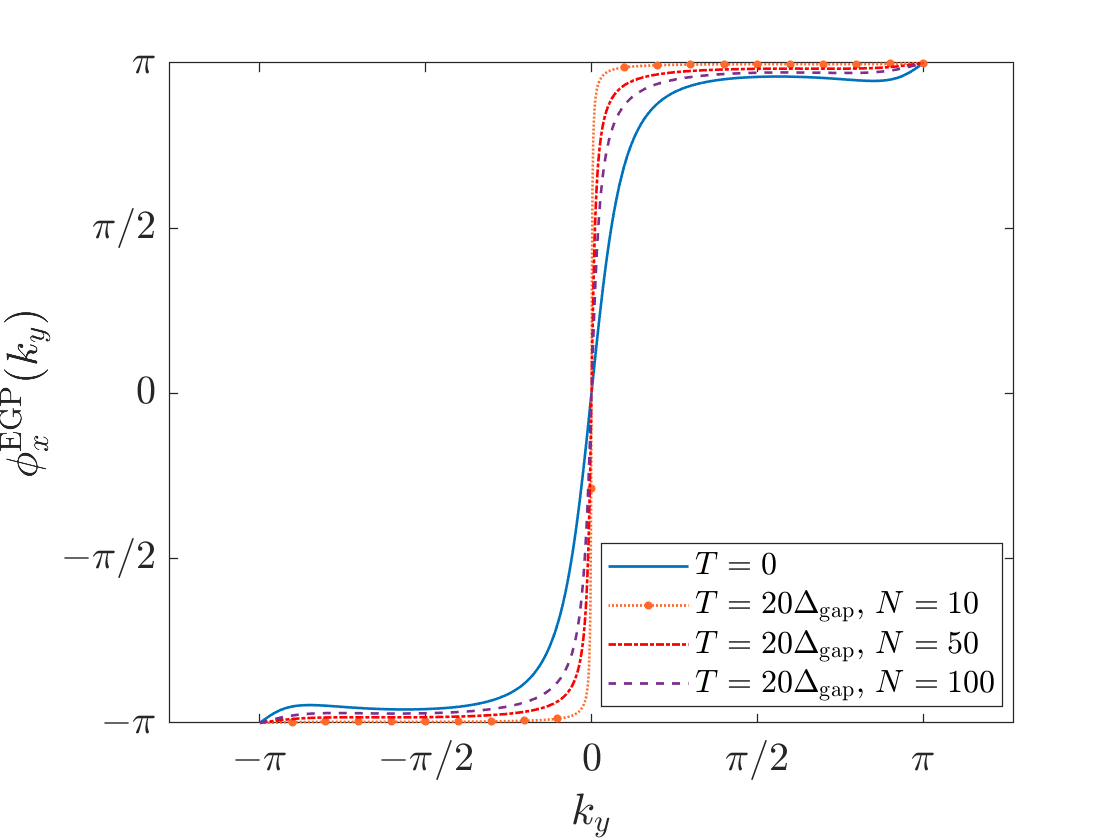}
	\includegraphics[width=0.8\columnwidth]{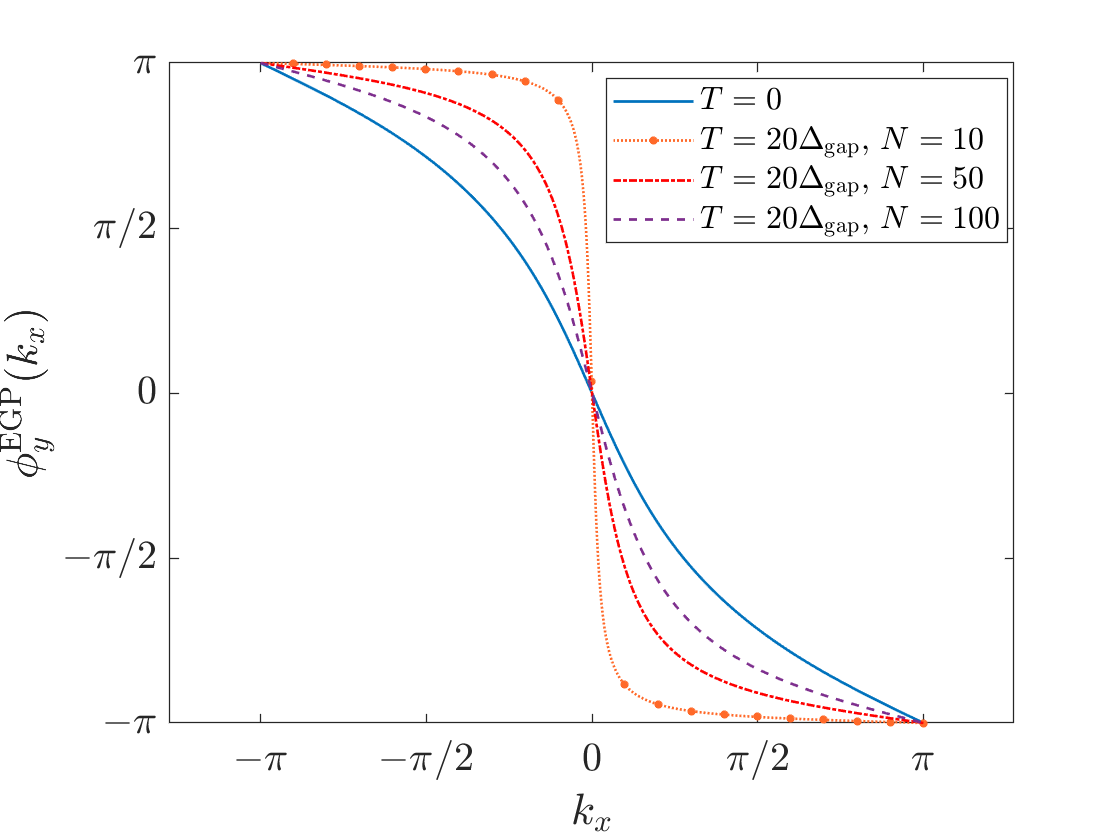}
	\end{center}
	\caption{EGP of Qi-Wu-Zhang model in
	$x$ direction as function of $k_y$ (top) and 
	in $y$ direction as function of $k_x$ (bottom) for different temperatures. One recognizes that in contrast to the Uhlmann case, the windings of both are exactly the same (with proper sign convention) for all temperatures even much above the single-particle gap.}
	\label{fig:QWZ-model}
\end{figure}

To illustrate our findings, we calculated the EGP components of the asymmetric Qi-Wu-Zhang model, eq.\eqref{fig:QWZ-model}, for a finite-temperature state for a finite system of $N\times N$ unit cells. The results are shown in Fig.\ref{fig:QWZ-model} for $T=0$ and a temperature much above the single-particle energy gap $T=20 \Delta_\textrm{gap}$. For the higher temperature we have also shown the results for different system sizes $N=10, 50, 100$. One clearly recognizes the gauge reduction discussed in \cite{Bardyn-PRX-2018}: Even at a temperature much above the single-particle gap the EGP approaches the ground-state value for increasing $N$. Its winding, which determines the topological invariant is furthermore independent of system size. Moreover, in contrast to the Uhlmann phase, the winding of both components $\phi^\textrm{EGP}_x$ and $-\phi^\textrm{EGP}_y$ is always the same for all finite temperatures $T<\infty$.

\section{summary}

We have shown that the ensemble geometric phase which 
has been used to define topological winding numbers for mixed states of one-dimensional, Gaussian fermion systems can straightforwardly be extended to two spatial dimensions and defines a Chern number if there is translational invariance. Different from approaches based on
other geometric phases for mixed states, such as the Uhlmann phase, this number is a true topological invariant as it is  
a two-dimensional integral over a proper Berry curvature. The latter is defined by the Bloch states of the fictitious Hamiltonian formed by the matrix of single-particle correlations in the Gaussian mixed state.  
Finite-temperature
states of non-interacting fermion models, which are  fully characterized by single-particle correlations
are Gaussian, but Gaussian states can also emerge as non-equilibrium steady states of systems coupled to specific Markovian reservoirs. In the first case the mixed-state Berry curvature is identical to that of the original Bloch Hamiltonian counting all bands below the chemical potential as long as $T<\infty$. For genuine non-equilibrium states no such correspondence exists and the fictitious Hamiltonian has a meaning of its own.
While the discussion in the present paper relies on the assumption of translational
invariance, allowing a mapping to decoupled one-dimensional chains
(see Fig.\ref{fig:decomposition}), we anticipate the results to hold also in the presence of disorder and with interactions.
This will  be subject of further studies.

\subsection*{acknowledgement}
Financial support from the DFG through SFB TR 185, project number 277625399  is gratefully acknowledged. 


\end{document}